\documentclass[iicol, pdflatex,sn-nature]{sn-jnl}% Math and Physical Sciences

\usepackage{mathrsfs}
\usepackage{graphicx}%
\usepackage{multirow}%
\usepackage{amsmath, amssymb, amsfonts}%
\usepackage{amsthm}%
\usepackage{mathrsfs}%
\usepackage[title]{appendix}%
\usepackage{xcolor}%
\usepackage{textcomp}%
\usepackage{manyfoot}%
\usepackage{booktabs}%
\usepackage{algorithm}%
\usepackage{algorithmicx}%
\usepackage{algpseudocode}%
\usepackage{listings}%
\usepackage{siunitx}
\usepackage{svg}
\usepackage{ulem}
\theoremstyle{thmstyleone}%
\theoremstyle{thmstyletwo}%

\theoremstyle{thmstylethree}%

\unnumbered% uncomment this for unnumbered level heads

\begin{document}
    %\title[The spin dynamics All-optical nonlinear Compton scattering]{The spin dynamics All-optical nonlinear Compton scattering}

    \title{Experimental Determination of Gamma-Ray Polarization in Strong-Field Nonlinear Compton Scattering}


\author[1]{\fnm{Pengpei} \sur{Xie}}\equalcont{These authors contributed equally to this work.}%\email{xiepengpei@stu.xjtu.edu.cn}
    \author[2]{\fnm{Mingyang} \sur{Zhu}}\equalcont{These authors contributed equally to this work.}%\email{consider\_zmy@sjtu.edu.cn}
    \author[2]{\fnm{Xichen} \sur{Hu}}%\email{xichenhu@sjtu.edu.cn}
 \author*[1]{\fnm{Yanfei} \sur{Li}}\email{liyanfei@xjtu.edu.cn}

    \author*[3]{\fnm{Yifei} \sur{Li}}\email{yflx@iphy.ac.cn}

    \author[1]{\fnm{Tianbing} \sur{Wang}}%\email{wangtianbing@stu.xjtu.edu.cn}

    \author[1]{\fnm{Bingjun} \sur{Li}}%\email{lbj\_xjtu@stu.xjtu.edu.cn}

    \author[2]{\fnm{Huitong} \sur{Zhai}}%\email{zhaihuitong@sjtu.edu.cn}
   \author[2]{\fnm{Bingzhan} \sur{Shi}}%\email{805466336@sjtu.edu.cn}

    \author[1]{\fnm{Zewei} \sur{Zhang}}%\email{vic\_0@stu.xjtu.edu.cn}
    \author[1]{\fnm{Ruiqi} \sur{Qin}}%\email{qrq2833757055@stu.xjtu.edu.cn}

        \author[2]{\fnm{Jie} \sur{Feng}}

    \author[3]{\fnm{Jinguang} \sur{Wang}}%\email{jgwang@iphy.ac.cn}
    \author[3]{\fnm{Xin} \sur{Lu}}%\email{luxin@iphy.ac.cn}

    \author[2]{\fnm{Liming} \sur{Chen}}%\email{lmchen@stu.xjtu.edu.cn}
    \author[3,4,5]{\fnm{Yutong} \sur{Li}}%\email{liyan@xjtu.edu.cn}

    \affil[1]{\orgdiv{Department of Nuclear Science and Technology, School of Energy and Power Engineering}, \orgname{Xi’an Jiaotong University}, \orgaddress{ \city{Xi'an}, \postcode{710049}, \country{China}}}

    \affil[3]{\orgdiv{Beijing National Laboratory for Condensed Matter Physics, Institute of Physics}, \orgname{Chinese Academy of Sciences}, \orgaddress{\city{Beijing}, \postcode{100190}, \country{China}}}

\affil[3]{\orgdiv{State Key Laboratory of Dark Matter Physics, Key Laboratory for Laser Plasmas (MoE), School of Physics and Astronomy}, \orgname{Shanghai Jiao Tong University}, \orgaddress{\city{Shanghai}, \postcode{200240}, \country{China}}}

\affil[4]{\orgdiv{School of Physical Sciences}, \orgname{University of Chinese Academy of Sciences}, \orgaddress{\city{Beijing},\postcode{1000490}, \country{China}}}

\affil[5]{\orgdiv{Attosecond Science Center}, \orgname{Songshan Lake Materials Laboratory}, \orgaddress{\city{Dongguan, Guangdong},\postcode{523808}, \country{China}}}

    \abstract{%Light–matter interactions driven by ultrahigh-intensity lasers have great potential to uncover the physics associated with quantum electrodynamics (QED) processes occurring in neutron stars and black holes. The nonlinear Compton scattering between an ultra-relativistic electron beam and an intense laser offers a direct probe into this regime. However, while spectral signatures of quantum radiation reaction have been observed, the spin-dependent polarization properties of the emitted radiation have remained experimentally elusive. Here we present the first experimental determination of the gamma-ray polarization state generated via nonlinear Compton scattering in the strong-field regime. In our experiment, electrons with energies of several hundred MeV interact with a plasma-mirror-reflected, ultrahigh-contrast laser pulse to emit a directed beam of gamma rays. By utilizing the azimuthal asymmetry of photoneutrons produced via the photodisintegration of deuterium in heavy water, we measure an average linear polarization degree of approximately 50\%. These results are corroborated by spin-resolved particle-in-cell simulations that benchmark the Locally Constant Field Approximation against the Locally Monochromatic Approximation. Our findings unambiguously rhttps://www.overleaf.com/project/690849214933628f0c2cac26#eveal the quantum and spin-dependent nature of the radiation process, paving the road to examine nonlinear Breit–Wheeler pair production and QED cascades.
   The polarization of gamma rays produced in strong-field quantum electrodynamics (SFQED) is a fundamental and long-standing prediction, the verification of which has remained elusive, limiting both foundational tests and applications. Here, we report the first experimental measurement of gamma-ray polarization generated via all-optical nonlinear Compton scattering. Colliding a laser-wakefield-accelerated electron beam with an intense counter-propagating laser pulse reflected from a plasma mirror, we produce bright gamma rays in the strong-field regime ($a_0 > 1$). For gamma rays with $a_0 \approx 3$, a linear polarization degree of $\sim 50\%$ is measured via the azimuthal asymmetry of photoneutrons from a deuterium target, and independently verified by a Compton polarimeter. 
   %This result is in agreement with strong-field QED calculations using the locally monochromatic approximation. 
   The results show excellent agreement with SFQED calculations employing the locally monochromatic approximation, while diverging from predictions based on the locally constant field approximation, highlighting the importance of quantum interference effects in this regime. Our work provides experimental evidence for polarization dynamics in SFQED, supports a key prediction of nonperturbative QED, and paves the way for compact, laser-driven sources of polarized gamma rays.}

    \maketitle
    \clearpage
    %\section{Introduction}
    \label{sec1}

    Strong-field quantum electrodynamics (SFQED), which describes light-matter interactions in extreme electromagnetic fields, opens a window to explore fundamental physics from astrophysical environments to future colliders \cite{gonoskov2022charged, Zhang2020,di2012extremely}.  While linear Compton scattering is known to produce highly polarized radiation, the transition into the nonlinear regime ($a_0 \gg 1$) opens a distinct and largely unexplored frontier. Here, theory predicts that the polarization state of emitted gamma rays becomes a highly sensitive observable, encoding rich dynamics of the interaction such as the effects of the electron’s instantaneous trajectory, potential spin correlations, and the order of multiphoton absorption \cite{Li2020Polarized, Blackburn_2021}. Experimental access to this polarization dimension is therefore essential to advance from spectral measurements to a complete understanding of nonperturbative QED dynamics, and to develop next-generation light sources with tailored, high-brightness polarization properties for applications in fundamental physics \cite{Howell_2022, BADELEK2004} and as polarized positron sources \cite{li2020production, Bambade2019ilc}.

    Accessing the core process of nonlinear Compton scattering (NCS) requires colliding ultrarelativistic electrons with ultra-intense laser pulses.  The interaction regime is governed by two fundamental dimensionless parameters.
    The first is the normalized laser amplitude $a_{0}\equiv |eA|/m_{e}c^{2}$ (where $e$ is the electron charge, $A$ is the laser vector potential, $m_{e}$ is the electron rest mass, and $c$ is the speed of light in vacuum).
    The condition $a_{0}\gg 1$ indicates a nonperturbative, strong-field interaction where electrons absorb multiple laser photons, leading to NCS.
    The second is the quantum nonlinearity parameter $\chi_{e}\approx \gamma E_{\perp}/E_{S}$ where $\gamma$ is the electron Lorentz factor, $E_{\perp}$ is the effective transverse field strength in the laboratory frame, and $E_{S}= m_{e}^{2}c^{3}/e\hbar \approx 1.32 \times 10^{18}$ V/m is the Schwinger critical field.
    The parameter $\chi_{e}$ quantifies the relative importance of quantum effects during photon emission.
    Tytically, $\chi_{e}\gtrsim 1$ denotes the strong quantum regime, where quantum radiation reaction and discrete photon emission dominate. Crucially, however, even when $\chi_{e}\ll 1$, provided  $a_{0}
    \gg 1$, distinct quantum effects such as the stochastic nature of photon emission can yield significant corrections to classical dynamics \cite{Hu2020,Neitz2013,gonoskov2022charged,ritus1985quantum, di2012extremely}.

    Experimental investigation of NCS has advanced in tandem with improvements
    in laser intensity.
    Foundational work was conducted at large-scale accelerator facilities, most notably the landmark SLAC experiment, where 46 GeV electrons
    interacted with a laser field of $a_{0}\approx 0.6$ to observe the absorption of up to four laser photons in the perturbative regime, confirming the basic multiphoton picture \cite{bula1996observation}.
    The advent of high-power,laser-wakefield acceleration (LWFA) enabled compact, all-optical investigations into the strong-field classical regime.
    A pivotal demonstration was the observation of high-order multiphoton absorption, in which a single hundreds-of-MeV electron ($\chi_e \ll 1$) was shown
    to coherently scatter over hundreds of laser photons ($a_{0}\sim 10$) into a
    single X-ray photon \cite{yan2017}. Subsequent experiments, employing
    higher-intensity lasers and GeV-class electron beams, progressed into the
    radiation-reaction regime, where unambiguous signatures of quantum recoil were
    identified in the emitted radiation spectra \cite{cole2018experimental,poder2018experimental}.
    More recently, such platforms have achieved NCS where a multi-GeV electrons absorbing
    over 400 photons to produce gamma rays with energies of several hundred MeV,
    corresponding to a $\chi_{e}\approx 0.46$, firmly placing the interaction within
    the SFQED regime where quantum effects are significant \cite{mirzaie2024all}.
    Despite significant progress in capturing spectral signatures of strong-field
    effects, a critical gap remains. That is, the polarization dynamics of gamma
    rays emitted via NCS, predicted to be a prominent feature in the nonlinear
    regime, have not yet been experimentally resolved. This gap impedes the full
    verification of SFQED polarization theory and the development of practical
    laser-driven polarized sources.

      \begin{figure*}[!htbp]
        \centering
        \includegraphics[width=0.95\textwidth]{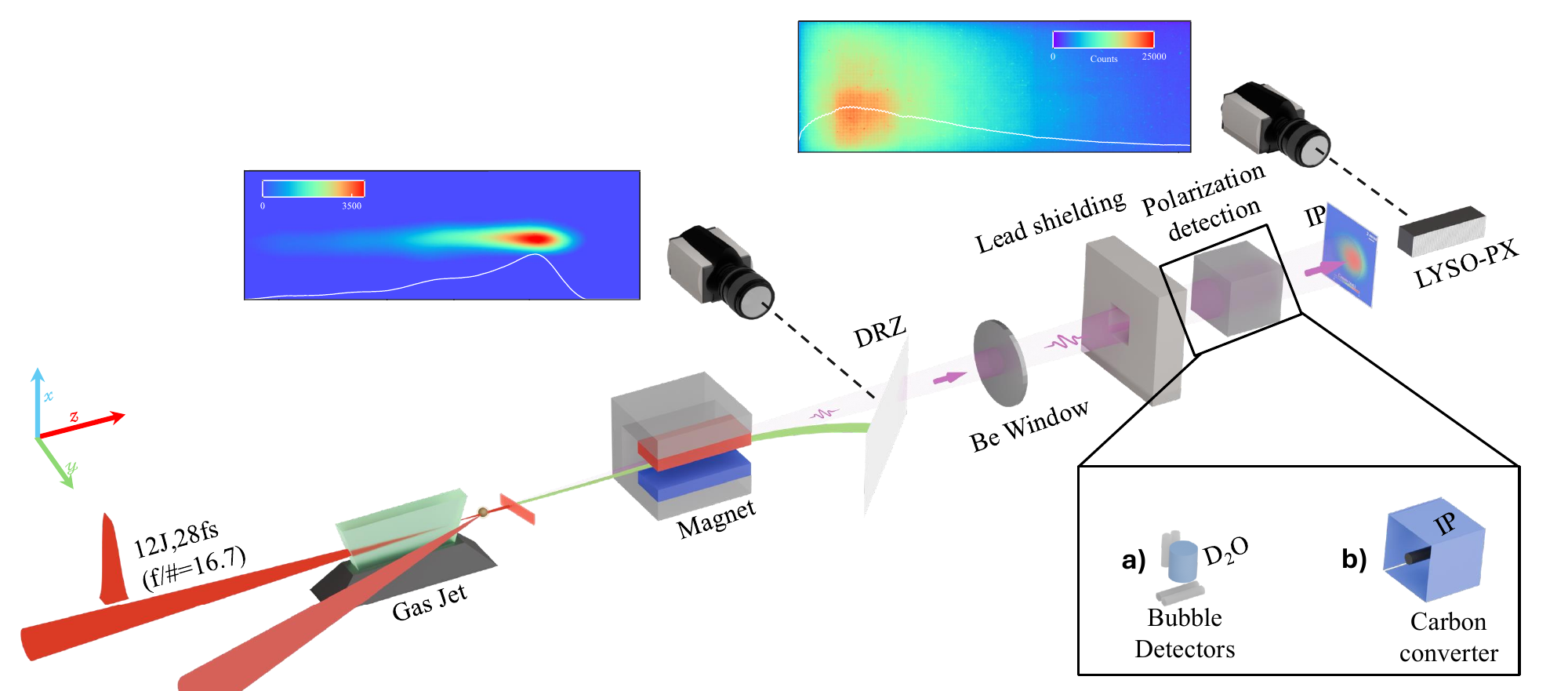}
        \smallskip
        \caption{{\normalcolor}\textbf{Experimental set-up}. Schematic of the all-optical experimental setup for studying polarized nonlinear Compton scattering.
A drive laser pulse generates an ultrarelativistic electron beam via LWFA in a gas jet. The residual laser is reflected and focused by a plasma mirror to collide with the electrons. The scattered electrons are analyzed by a magnetic spectrometer. The emitted gamma rays are characterized using an Imaging Plate (IP) for beam profile and a pixelated LYSO (LYSO-PX) calorimeter for spectrum. Their polarization is determined via two interchangeable detection modules (bottom insets): a) a heavy-water ($\mathrm{D_2O}$) target monitored by two bubble detectors positioned parallel (side) and perpendicular (bottom) to the horizontal linear polarization of the incident laser, and b) a solid carbon converter for Compton scattering asymmetry analysis using IPs.
%Schematic  of the all-optical nonlinear Compton scattering experiment with a plasma mirror (CPM). Electrons are first accelerated  via LWFA through a PW laser pulse interacting with a gas jet to hundreds of MeV, then collide with the ultra-intense laser field reflected by the CPM with a colliding angle of 30$^\circ$.  Gamma rays with energies up to 60 MeV are generated during the interaction via  NCS. The energy spectrum of scattered electrons  are detected with an electron spectrometer with a 1.33-T dipole magnet. The gamma rays generated are characterized by a scintillating crystals of LYSO—a 5-mm-thick, 90-mm-diameter, single-crystal LYSO, for imaging the beam profile, and a pixelated one, LYSO-PX, for measuring the gamma-ray spectrum. The gamma-ray polarization is diagnosed through the photodisintegration of deuterium in a heavy water target, with emitted neutrons detected by an array of scintillation detectors positioned around the target and through a solid conversion target detected by image planes (IPs).
}
        \label{fig:fig1}
    \end{figure*}

    To address this, we present the first experimental study that resolves the polarization of gamma rays from all-optical NCS. Using a compact plasma-mirror (PM) setup, we collide a LWFA electron beam with an intense, reflected laser pulse to produce gamma rays in the strong-field regime. We measure an average linear polarization degree of approximately 50\% for gamma rays emitted in NCS events with $a_{0}\approx 3$. Our findings, supported by strong-field QED calculations based on the Locally Monochromatic Approximation \cite{Blackburn_2021, tangFullyPolarizedCompton2024}, validate this key dynamical prediction. This work extends the experimental study of laser-driven NCS beyond intensity and spectral measurements to access the polarization degree of freedom, providing experimental evidence for polarization dynamics in the nonlinear regime and demonstrating a compact, laser-driven source of polarized gamma rays via NCS.

    %Using a compact plasma-mirror setup at the SECUF (Synergetic Extreme Condition User Facility) facility, we collide LWFA-generated electron beams with intense, reflected laser pulses, leading to the emission of gamma rays. Through deuterium photodisintegration, we measure an average linear polarization degree of approximately 50\% for gamma rays emitted in NCS events with $a_{0}\sim 5$. Our findings, supported by calculations based on the Locally Monochromatic Approximation  \cite{Blackburn_2021, tangFullyPolarizedCompton2024}, validate a key dynamical aspect of the theory.  This work shifts the study of laser-driven NCS from intensity and spectrum alone to include the complete polarization observable, establishing a new diagnostic capability for extreme light-matter interactions and a practical route to compact, high-brightness polarized gamma-ray sources.

  Our all-optical experiment employs a compact, single-beam geometry designed for inherent spatiotemporal synchronization at the SECUF (Synergetic Extreme Condition User Facility) facility. As illustrated in Fig. \ref{fig:fig1}, a single petawatt-class laser pulse first drives LWFA in a gas target, generating a multi-hundred-MeV electron beam. The residual laser pulse is reflected and focused by  plasma mirror positioned 15° off-normal, generating an ultra-intense field for the collision with the electron beam% \cite{chen2025enhanced}
  . This self-aligned setup eliminates the need for a separate, jitter-sensitive scattering beamline, a major technical hurdle in conventional two-beam experiments.

  The scattered electrons are characterized by a magnetic spectrometer. The emitted gamma rays are diagnosed using a layered approach: their spatial profile is captured by an imaging plate (IP), and their energy spectrum is resolved with a pixelated LYSO (LYSO-PX) calorimeter. Polarimetry is performed via two complementary, self-consistent techniques to ensure robustness. The primary method utilizes the deuteron photodisintegration reaction in a heavy-water target\cite{yan2019precision}, where the azimuthal anisotropy of emitted neutrons encodes the polarization information, detected by a surrounding array of bubble detectors. In parallel, a Compton polarimeter, employing a solid conversion target and IPs positioned parallel and perpendicular to the polarization axis to measure the scattering asymmetry, provides an independent polarization measurement.

  %To isolate the pure NCS signal from the bremsstrahlung background generated as electrons pass through the PM substrate, we developed a spectral retrieval algorithm. It decomposes the composite detector response into simulated NCS and bremsstrahlung components, enabling the simultaneous extraction of the deconvolved gamma-ray spectrum and the effective laser intensity $a_0$ at the interaction point (see Methods for analysis details).

    \begin{figure*}[htbp]
        \centering
        \includegraphics[width=0.99\linewidth]{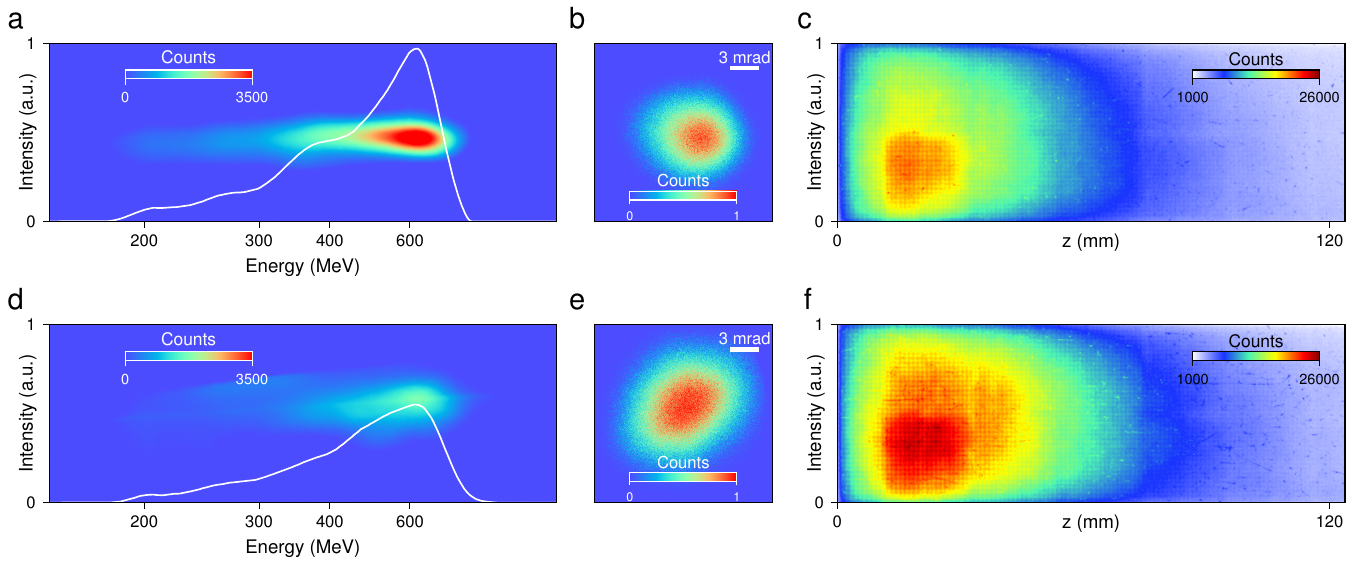}
      \caption{\textbf{Distinguishing Compton scattering from bremsstrahlung background.}
\textbf{a-f,} Single-shot measurements under two conditions: bremsstrahlung-only (\textbf{a-c}) and bremsstrahlung with NCS (\textbf{d-f}).
\textbf{a, d,} Electron energy spectra (false-colour DRZ images with unified scale). White curves: lineouts showing the energy distribution.
\textbf{b, e,} Gamma-ray beam spatial profiles on an IP.
\textbf{c, f,} Gamma-ray spectral signals recorded by the LYSO-PX calorimeter. The white curves represent the longitudinal signal profiles obtained by vertical integration.
To facilitate direct comparison of signal yields, these profiles are normalized to the global maximum within each measurement modality: the electron spectra (\textbf{a} and \textbf{d}) are normalized to the peak intensity of \textbf{a}, while the calorimeter signals (\textbf{c} and \textbf{f}) are normalized to the peak intensity of \textbf{f}.
}
        \label{fig2}
    \end{figure*}

To confirm the occurrence of laser-driven Compton scattering and distinguish its signal from the ever-present bremsstrahlung background, we performed a controlled, pairwise comparison. Two comparative shots were obtained under identical experimental conditions, distinguished solely by the longitudinal position of the PM relative to the gas jet nozzle: one with the PM placed 10 mm downstream to effectively suppress the laser-electron collision (bremsstrahlung reference; Fig. \ref{fig2}a-c), and a second with the PM positioned 2 mm downstream to enable collision (bremsstrahlung plus NCS; Fig. \ref{fig2}d-f).

A direct quantitative comparison of the electron spectra (Fig. \ref{fig2}a, d) reveals that the integrated charge in the collision shot is reduced to 65.1\% of the bremsstrahlung-only reference. Specifically, the spectral intensity at the 600 MeV peak drops to 54.3\% of the reference level, while the cutoff energy remains identical.
Despite this reduction in electron flux, the accompanying gamma-ray signal is dramatically enhanced. The spatial profile of the gamma-ray beam recorded on an IP (Fig. \ref{fig2}e), when normalized to its peak intensity, exhibits a measurably larger divergence compared to the pure bremsstrahlung case (Fig. \ref{fig2}b). This subtle spatial signature is strongly corroborated by the energy-resolved signal in the LYSO-PX calorimeter (Fig. \ref{fig2}c, f). With the laser present, the total integrated gamma-ray yield increases by a factor greater than 1.25, and the maximum deposited energy extends to significantly higher values. The fact that a weaker electron beam produces a far brighter and more energetic gamma-ray signal demonstrates the activation of an efficient, laser-driven Compton scattering mechanism.

\begin{figure*}[!htbp]
    \centering
    \includegraphics[width=0.95\linewidth]{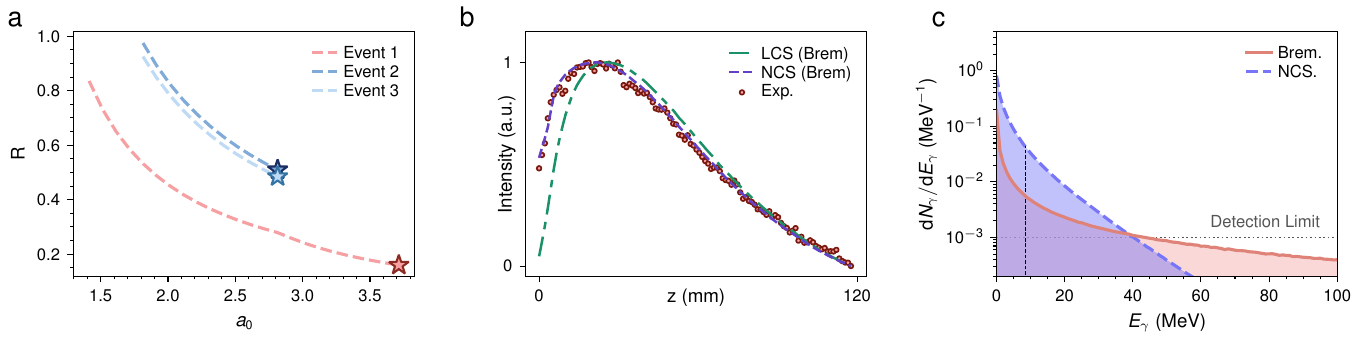}
  \caption{\textbf{Spectral decomposition proving nonlinear Compton scattering.}
\textbf{a,} Inferred effective laser intensity $a_0$ and electron participant fraction $R$ (defined as the ratio of electrons undergoing Compton scattering to those producing bremsstrahlung) for three independent interaction events (labeled Event 1--3). The stars mark the best-fit parameters derived from the maximum likelihood analysis.
\textbf{b,} Signal reconstruction for Event 1. The experimental LYSO profile (red circles) is compared with best-fit models assuming nonlinear (NCS, purple dashed line) or linear (LCS, green dashed-dotted line) Compton scattering superimposed on the bremsstrahlung background.
\textbf{c,} Deconvolved gamma-ray spectra for Event 1. The extracted NCS component (blue dashed line) extends well beyond the linear cutoff energy (vertical black dashed line, $\approx 8.4$ MeV), while the red curve depicts the bremsstrahlung background.}
    \label{fig4}
\end{figure*}

%As shown in Fig.~\ref{fig3}, the experimental data for \(\mathcal{S}_{\mathrm{NCS}}\) are well-fitted by this function (solid line). The normalized yield is highest at the optimal overlap position and decreases monotonically as the mirror is retracted and \(a_0\) diminishes. The fitted Rayleigh length \(z_R\approx 1.50\) mm  is larger than the ideal value calculated from the initial laser parameters ($\sim 0.48$ mm). This is attributed to several experimental factors: (i) the convolution of the finite-sized electron beam with the laser intensity profile, which broadens the measured spatial response; (ii) wavefront distortion introduced by the plasma mirror ; and (iii) modification of the laser’s spatiotemporal properties during the interaction. The latter includes nonlinear frequency upshifts and spectral broadening as the pulse propagates through the underdense plasma and reflects from the overdense plasma mirror, which alter the effective wavelength and beam caustic. Despite these complexities in precisely determining the absolute laser profile at the interaction point, the strong functional dependence of \(\mathcal{S}_{\mathrm{NCS}}\) on PM displacement and its vanishing without overlap provide unambiguous evidence for the laser-driven origin of the signal.

%\sout{Having established that the gamma-ray signal is generated by the laser-driven Compton process and scales with the laser intensity $a_0$, we next determine whether this scattering occurs in the linear ($a_0 \ll 1$) or nonlinear ($a_0 > 1$) regime in Fig.\ref{fig4}.}
Having confirmed the laser-driven origin of the gamma-ray signal, our next critical step is to quantitatively extract the effective laser intensity ($a_0$) and definitively determine whether the scattering occurs in the perturbative linear ($a_0 \ll 1$) or strong-field nonlinear ($a_0 > 1$) regime. A systematic scan of the plasma mirror position (Supplementary Fig. S1) identified $z = 1$ mm as the optimal location maximizing the gamma-ray yield. From six consecutive high-signal shots acquired at this position, three were selected for detailed spectral analysis to extract $a_0$, while the remaining three shots under the same optimal conditions were used for subsequent polarization measurements.
%This analysis focuses on three consecutive high-signal shots acquired at the optimal PM position of $z$=1 mm, (the scan of PM position Supplementary Fig. S1), which were subsequently used for polarization measurements. 
%(Detailed raw signals for the LYSO array at various spatial positions, background subtraction procedures, and further details on the spectral decomposition are provided in the Methods and Supplementary Fig. S2.) 

To extract the pure NCS spectral component and the effective laser intensity $a_0$ from the composite signal, we developed a precise spectral decomposition framework. For each shot, the bremsstrahlung contribution was simulated using GEANT4, with the measured electron spectrum, whose energy loss when traversing the plasma mirror substrate is validated to be negligible (Supplementary Fig. S3), as the input source. The NCS contribution was modeled using the Locally Monochromatic Approximation (LMA), which provides the gamma-ray spectrum for a given $a_0$. The combined simulated response (bremsstrahlung and NCS) on the LYSO-PX was then compared to the experimental signal. By varying $a_0$ and the fraction $R$ of electrons participating in NCS, we performed a maximum-likelihood fit to determine the most probable parameter pair for each shot (see Methods).

The resulting values of the fitted laser intensity $a_0$ for the three shots are 2.8, 2.8, and 3.7 (Fig.~\ref{fig4}a), with a median of 3, which is in good agreement with 3D-PIC simulations (Supplementary Fig. S5). All values exceed unity, providing  evidence that the interactions occurred in the nonlinear regime.

The critical model comparison is shown for a representative shot in Fig.~\ref{fig4}b. The experimental LYSO-PX signal (red point) is excellently reproduced by the combined bremsstrahlung and NCS (LMA) model (perple dashed line). In stark contrast, a model assuming linear Compton scattering (LCS, green dashed-dotted line), even under the unrealistic assumption that 100\% of electrons undergo LCS, fails to match the signal’s amplitude and shape. This discrepancy rules out a linear scattering process as the source of the bright gamma-ray signal.

Another evidence for nonlinear scattering is provided by the deconvolved gamma-ray spectra. Figure \ref{fig4}c presents the extracted NCS spectral component (blue dashed line) for Event 1. The retrieved spectrum extends to energies significantly beyond the maximum kinematic energy attainable via linear Compton scattering ($\approx 8.4$ MeV, indicated by the vertical black dashed line). This signal, persisting well above the detection limit, provides unambiguous evidence of multiphoton absorption, conclusively establishing the nonlinear nature of the interaction.

%Having firmly established the nonlinear nature of the generated gamma rays, we now focus on the central result of this study: the single-shot determination  of their linear polarization state. {\color{red} Fig. [4]\normalcolor}presents the azimuthal distribution of photoneutrons recorded in a representative single-shot event. A distinct asymmetry is clearly visible, yielding a raw measured modulation factor of $\mathcal{A}_{raw}\approx 0.28$. However, it is crucial to recognize that this measured signal comprises both  the highly polarized nonlinear Compton gamma rays and the unpolarized bremsstrahlung background generated by the plasma mirror substrate, which effectively  dilutes the observed asymmetry. To extract the intrinsic polarization of the  NCS source, we applied a correction based on the signal-to-background ratio derived from our Geant4 simulations. After decoupling the contribution of the unpolarized bremsstrahlung, the corrected asymmetry corresponds to a high intrinsic linear polarization degree of $P_{\gamma}\approx 50\%$. This result is in excellent agreement with the predictions of our spin-resolved LMA simulations, confirming the effective transfer of laser spin angular momentum to the emitted gamma rays.

\begin{figure*}
    \centering
    \includegraphics[width=0.95\linewidth]{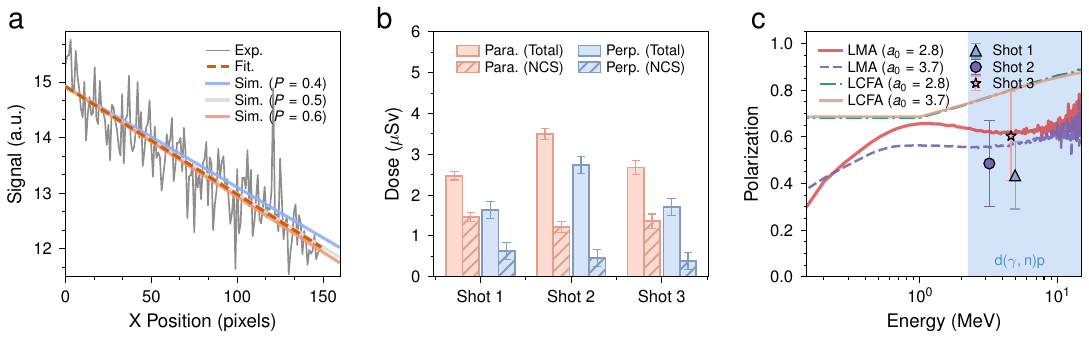}
    \caption{\textbf{Determination and validation of gamma-ray polarization.}
\textbf{a,} Comparison of the experimentally measured spatial profile of the gamma-ray signal (integrated IP signal, grey line) with Geant4 simulations (colored solid lines, assuming a known gamma-ray spectrum with varying polarization degrees of  0.4, 0.5, 0.6). \textbf{b,} Neutron doses measured by bubble detectors positioned parallel (Para., red) and perpendicular (Perp., blue) to the laser polarization axis. The solid bars represent the total measured dose (signal + background), while the hatched bars denote the isolated NCS component obtained after explicitly subtracting the bremsstrahlung background. Error bars represent the standard deviation. \textbf{c,} Validation of the retrieved polarization. The plot displays the experimental polarization degree (derived from the neutron yield asymmetry) and average energy for three shots, compared against theoretical predictions based on the LCFA and LMA models for the emitted photon polarization.}
    \label{fig5}
\end{figure*}

With the nonlinear nature of the Compton scattering established, we now present the main result of this work: the first experimental measurement of the polarization of gamma rays generated via all-optical NCS. To characterize the polarization across the full spectrum, we employed two complementary diagnostics separated by the deuterium photodisintegration threshold ($\approx 2.2$ MeV). While the high-energy component ($>2.2$ MeV) was monitored using the heavy water setup, the polarization of low-energy photons  was mainly determined via Compton scattering asymmetry using a solid Carbon converter. Linearly polarized gamma rays undergo Compton scattering with a cross-section dependent on the azimuthal angle $\phi$ relative to the polarization vector, resulting in an anisotropic spatial distribution of scattered photons. This distribution was recorded on an IP stack, yielding the intensity profile shown in Fig.~\ref{fig5}a (grey line). By comparing the measured signal gradient with $\mathrm{GEANT4}$ simulations for varying polarization degrees (solid colored lines, see Supplementary Fig. S6 for the simulated azimuthal profiles), we infer an average polarization of $P = 0.5$ for this lower-energy component.

To probe the polarization of high-energy photons, we employed deuteron photodisintegration using a heavy-water ($\mathrm{D_2O}$) target. The $d(\gamma,n)p$ reaction possesses a strong analyzing power, yielding an azimuthal neutron anisotropy that directly encodes the incident gamma-ray polarization. Neutrons were detected by bubble detectors positioned parallel (Para.) and perpendicular (Perp.) to the laser polarization axis. Figure~\ref{fig5}b presents the dosimetry results for three independent shots (raw bubble detector data and deconvolved incident photon spectra are provided in Supplementary Fig. S7). The solid bars represent the total measured doses (including the bremsstrahlung background), whereas the hatched bars denote the isolated NCS signals obtained after background subtraction. Crucially, the extracted NCS component exhibits a pronounced asymmetry, with the parallel dose consistently exceeding the perpendicular dose, enabling the robust extraction of the polarization degree.

The retrieved polarization degrees and corresponding mean photon energies are summarized in Figure~\ref{fig5}c. The analysis yields mean polarization values of $P \approx 43.5\%$ at $\langle E_\gamma \rangle \approx 4.9$ MeV (Shot 1), $P \approx 48.6\%$ at $\langle E_\gamma \rangle \approx 3.2$ MeV (Shot 2), and $P \approx 60.3\%$ at $\langle E_\gamma \rangle \approx 4.6$ MeV (Shot 3). These experimental results are plotted against theoretical predictions from two complementary strong-field QED frameworks. Overlaid are calculations using the LMA and the locally constant field approximation (LCFA) for representative intensities of $a_0 = 2.8$ and $a_0 = 3.7$, which span the range of experimental conditions.

The experimental results exhibit excellent agreement with the LMA calculations across all three shots. In contrast, the LCFA systematically overestimates the polarization degree, with deviations becoming increasingly pronounced at lower photon energies. This discrepancy is not merely quantitative but reflects a fundamental limitation of the LCFA in the intermediate-intensity regime ($a_0 \sim 3$). The physical origin of this deviation is now well established. The LCFA, while computationally efficient and widely adopted in particle-in-cell simulations, relies on the assumption that the formation length of the emission process is much shorter than the scale of field variation. In the regime where $a_0 \sim \mathcal{O}(1)$ to $\mathcal{O}(10)$, however, this assumption breaks down: the formation length becomes comparable to the laser wavelength, and quantum interference effects arising from the finite spatiotemporal extent of the interaction can no longer be neglected. These omitted effects include the correct low-energy asymptotics of the photon spectrum, the preservation of harmonic structure, and spin-dependent contributions to the emission probability, all of which influence the resulting polarization. The LMA, by contrast, was specifically formulated to address these limitations by treating the fast carrier-frequency oscillations exactly while approximating the slowly varying envelope, thereby preserving the multiphoton structure and interference effects that govern polarization transfer in the transition from perturbative to non-perturbative QED. Indeed, full QED calculations in pulsed plane-wave backgrounds have recently demonstrated that the LMA achieves quantitative agreement with the exact theory across the entire intensity spectrum, whereas the LCFA exhibits systematic deviations that worsen with decreasing intensity~\cite{tangFullyPolarizedCompton2024}. Our experimental results are in full agreement with this theoretical benchmark, providing the first experimental confirmation that the LMA correctly captures the polarization dynamics in the intermediate regime.

The consistency between the $\sim 50\%$ polarization extracted from the low-energy Compton channel (Fig.~\ref{fig5}a) and the high-energy photodisintegration measurements presented here confirms that all-optical nonlinear Compton scattering produces highly polarized gamma rays across the full emitted spectrum, a foundational prediction of strong-field QED. More importantly, these data establish polarization as a uniquely sensitive observable for discriminating between competing theoretical approaches. While previous milestone experiments inferred quantum signatures from macroscopic observables such as energy loss and spectral broadening, polarization probes the microscopic spin-dependent emission probabilities with far greater specificity. The clear failure of the LCFA and the quantitative success of the LMA in reproducing our data demonstrate that interference effects and the finite scale of the formation region cannot be neglected in this regime, establishing polarization as a critical benchmark for strong-field QED theories.. This validation establishes the LMA as a reliable tool for designing next-generation polarized gamma-ray sources and provides a rigorous foundation for interpreting higher-order SFQED processes, such as vacuum birefringence and nonlinear Breit-Wheeler pair production, where polarization will serve as the definitive signature.

In summary, we have performed the first experimental measurement of gamma-ray polarization in all-optical nonlinear Compton scattering. By colliding a laser-wakefield-accelerated electron beam with an intense counter-propagating pulse ($a_0 \sim 3$), we generated bright gamma rays with linear polarization degrees of $43.5\%$ to $60.3\%$ for mean photon energies between $3.2$ and $4.9$ MeV, measured via complementary deuterium photodisintegration and Compton polarimetry. The excellent agreement with LMA-based strong-field QED calculations provides the first experimental validation of polarization dynamics in the nonlinear regime, confirming a long-standing prediction of non-perturbative QED and establishing the LMA as a reliable tool for designing future polarized gamma-ray sources. This work paves the way for compact, laser-driven polarized gamma-ray sources and provides a rigorous foundation for probing higher-order QED processes, such as vacuum birefringence and nonlinear Breit-Wheeler pair production, where polarization serves as the definitive signature.

\section{Acknowledgments}

    We acknowledge funding from the National Natural Science Foundation of China (Grant Nos. 12222507, 12335016, W2412039, and U24A2016) and the Scientific Research Innovation Capability Support Project for Young Faculty (Grant No. ZYGXQNJSKYCXNLZCXM-E12). This work was supported by the Synergetic Extreme Condition User Facility (SECUF, https://cstr.cn/31123.02.SECUF.D3).

    \section{Methods}
    \label{sec11}

    \subsection{Laser wakefield acceleration }
The experiment was performed at the Ultrafast X-ray dynamics experimental station of the Synergetic Extreme Condition User Facility (SECUF), Institute of Physics, Chinese Academy of Sciences. A linearly polarized 800-nm laser pulse with a duration of 28 fs was focused onto a 6-mm-long supersonic gas nozzle using an $f/16.7$ off-axis parabolic mirror ($f = 3$ m). Delivering 12 J of energy to the target, the laser was focused to a near-Gaussian spot with a radius of $w_{0}=11\,\mu$m. This yielded a peak intensity of $4.3\times10^{19}\,\text{W/cm}^{2}$, corresponding to a normalized vector potential of $a_{0}\approx 6$. The gas target consisted of a mixture of 99$\%$ He and 1$\%$ O$_{2}$, resulting in a plasma density of $2.5\times10^{18}\,\text{cm}^{-3}$. Following the interaction, the electrons and radiation propagated through a 170-cm vacuum section before exiting into the ambient atmosphere via a 380-$\mu$m-thick beryllium window.

    \subsection{Electron beam diagnostics}

    The electron energy spectra were characterized using a magnetic spectrometer consisting of a 0.98-T dipole magnet ($16\,\text{cm} \times 8\,\text{cm}$) with its entrance positioned 50 cm downstream from the target. The dispersed electrons were detected on a DRZ-High scintillator screen ($\mathrm{Gd_{2}O_{2}S:Tb}$) located 118 cm downstream of the magnet exit. For absolute charge calibration, an Imaging Plate (IP) was placed directly behind the scintillator screen. To monitor the electron beam pointing and divergence, a retractable auxiliary DRZ screen was installed at the magnet entrance; this screen was removed from the beam path during radiation measurements to minimize bremsstrahlung background. The spectrometer covered an energy range of 100--1200 MeV, achieving an energy resolution of $\sim2$ MeV at 400 MeV.

    \subsection{Radiation diagnostics}

The generated gamma-ray beam exited the vacuum chamber through a beryllium window. The beam spatial profile was recorded by an IP positioned 80 cm downstream of the beryllium window. For energy spectral measurements, a pixelated Cerium-doped Lutetium Yttrium Orthosilicate (Ce:LYSO) calorimeter was placed 121 cm downstream of the IP. The IP served as a transmission detector to monitor the beam profile, and its material budget was explicitly included in the subsequent detector simulations for the calorimeter spectra retrieval.

The calorimeter consists of scintillator elements ($1 \times 1 \times 40\,\text{mm}^{3}$) arranged in an orthogonally alternating pattern, separated by $0.1$-mm-thick barium sulfate ($\text{BaSO}_{4}$) reflective layers to minimize optical crosstalk. The scintillation light produced by electromagnetic showers was collected by a 3-inch lens positioned $18\,\text{cm}$ from the detector surface and imaged onto a 16-bit Electron-Multiplying Charge-Coupled Device (EMCCD) camera.

To accurately retrieve the gamma-ray energy spectra from the raw EMCCD images, we developed a Geant4-assisted forward-fitting algorithm. The bremsstrahlung background, originating from electrons traversing the plasma mirror substrate, was explicitly modeled by inputting the experimentally measured electron spectra into a Geant4 simulation of the full detector setup. Concurrently, a library of theoretical NCS photon spectra was generated using the Locally Monochromatic Approximation (LMA) across a grid of laser intensities. The detector response to these theoretical spectra was simulated to construct a response matrix. The experimental energy deposition data were then fitted using a linear combination of the simulated bremsstrahlung background and the NCS response library, allowing for the simultaneous extraction of the effective laser intensity ($a_{0}$) and the deconvolved gamma-ray spectrum.

    \subsection{Polarization diagnostics}

    To characterize the polarization across the full spectral range, two complementary and interchangeable polarimetry configurations were employed.

For high-energy gamma rays, the linear polarization was determined using a photoneutron polarimeter based on the deuteron photodisintegration reaction $d(\gamma,n)p$. For these measurements, the LYSO calorimeter was removed, and a cylindrical polyethylene container (diameter 6.6 cm, length 7.6 cm, wall thickness 3 mm) filled with heavy water ($\mathrm{D_{2}O}$) was positioned on the beam axis, 255 cm downstream of the source. The emitted photoneutrons were detected by two pairs of bubble detectors (BTI-PND, Bubble Technology Industries) arranged azimuthally around the target at angles of $0^{\circ}$ (horizontal, parallel to the laser polarization) and $90^{\circ}$ (vertical, perpendicular to the laser polarization). Each detector pair consisted of two units with distinct sensitivities (nominally 2.7 and 4.5 bubbles/$\mu$Sv) to optimize the dynamic range. These detectors were selected for their insensitivity to the intense gamma-ray flash and high efficiency for fast neutrons.

The measurement relies on the strong azimuthal anisotropy of photoneutron emission, which peaks perpendicular to the polarization vector of the incident gamma rays. The azimuthal asymmetry, $\mathcal{A}_{\text{raw}}$, was calculated from the contrast in bubble counts between the vertical and horizontal detector pairs. To retrieve the intrinsic polarization degree, this raw asymmetry was corrected using the analyzing power of the detection system, which was determined via comprehensive Geant4 simulations accounting for the specific target geometry, neutron transport, and the dilution effect from the unpolarized bremsstrahlung background.

For the lower-energy component, polarization was measured via Compton scattering asymmetry. The $\mathrm{D_{2}O}$ target was replaced by a solid carbon converter block. The spatial distribution of scattered photons was recorded by an IP stack positioned downstream. Linearly polarized gamma rays undergo Compton scattering with a differential cross-section dependent on the azimuthal angle relative to the polarization vector. The degree of polarization was extracted by comparing the azimuthal intensity modulation of the recorded signal with Geant4 simulations of the scattering process for varying polarization degrees.

\subsection{Data analysis for the normalized NCS yield}
The derivation and extraction of the normalized yield $\mathcal{S}_{\mathrm{NCS}}$, as presented in Fig. 3, are detailed below.
\textbf{}
\begin{enumerate}
    \item \textbf{Theoretical scaling and definition.}
    The definition of $\mathcal{S}_{\mathrm{NCS}}$ is grounded in the theoretical scaling of nonlinear Compton scattering. In the nonlinear regime ($a_0 > 1$), the total energy radiated by a single electron, $W_{\text{rad}}$, can be evaluated as the product of the number of emitted photons and their average energy. This follows from the approximate number of emitted photons, $N_\gamma \approx 5 \alpha \hbar \omega_0 a_0 \tau/(\sqrt{3}m_e c^4)$ \cite{Thomas2012}, and the average photon energy, $\overline{\omega}_\gamma \approx E_e \chi_e$, where the electron quantum parameter is $\chi_e = (2\hbar/(m_e^2 c^4)) \omega_0 E_e a_0$. Therefore, the total radiated energy by a single electron scales as $W_{\text{rad}} \approx N_\gamma \overline{\omega}_\gamma \propto a_0^2 E_e^2$.Consequently, extending this to an electron beam with total charge $Q_e$ and mean square energy $\langle E_e^2 \rangle$, the integrated NCS signal deposited in the calorimeter is expected to scale as $\Sigma_{\mathrm{NCS}} \propto a_0^2 Q_e \langle E_e^2 \rangle$. To decouple the laser intensity dependence from electron beam fluctuations, we define the normalized yield as:
    \[
    \mathcal{S}_{\mathrm{NCS}} \equiv \frac{\Sigma_{\mathrm{NCS}}}{Q_e \cdot \langle E_e^2 \rangle}.
    \]
    By definition, $\mathcal{S}_{\mathrm{NCS}}$ serves as a direct proxy for the squared laser intensity, $\mathcal{S}_{\mathrm{NCS}} \propto a_0^2$.

    \item \textbf{Bremsstrahlung background subtraction.}
    The total signal recorded by the calorimeter, $\Sigma_{\mathrm{tot}}$, comprises contributions from both NCS and bremsstrahlung. To isolate the NCS component $\Sigma_{\mathrm{NCS}}$, the bremsstrahlung background was characterized using ``null-collision'' shots where the plasma mirror was displaced by $>5\,\mathrm{mm}$ to prevent laser-electron overlap. For these reference shots, the signal exhibited a strictly linear proportionality to the product of charge and mean energy, consistent with thin-target bremsstrahlung emission:
    \[
    \Sigma_{\mathrm{brems}} = \alpha (Q_e \cdot \langle E_e \rangle) + C.
    \]
    The proportionality coefficient $\alpha$ was determined via a linear fit forced through the origin . For each interaction shot, the NCS signal was extracted by subtracting the expected bremsstrahlung contribution calculated from the measured electron parameters: $\Sigma_{\mathrm{NCS}} = \Sigma_{\mathrm{tot}} - \Sigma_{\mathrm{brems}}$.

    \item \textbf{Spatial profile fitting.}
    The longitudinal variation of the normalized yield $\mathcal{S}_{\mathrm{NCS}}(z)$ traces the on-axis evolution of the laser intensity $a_0^2(z)$. Assuming a Gaussian focal spot, the intensity follows a Lorentzian profile:
    \[
    a_0^2(z) \propto \left[1 + \frac{(z - z_0)^2}{z_R^2}\right]^{-1},
    \]
    where $z_0$ is the focal position and $z_R$ is the Rayleigh length. The experimental data were fitted with the corresponding zero-baseline Lorentzian function\cite{Salamin2022}:
    \[
    \mathcal{S}_{\mathrm{NCS}}(z) = A \left[1 + \frac{(z - z_0)^2}{z_R^2}\right]^{-1}.
    \]
    The best-fit parameters ($A$, $z_0$, $z_R$) were extracted using a nonlinear least-squares algorithm and are provided in Fig. 3.
\end{enumerate}

\subsection*{Spectral decomposition and extraction of $a_0$}
The extraction of the NCS spectral content and the effective laser intensity $a_0$ was achieved through a physics-based, forward-fitting decomposition of the composite LYSO-PX calorimeter signal. The analysis procedure for each shot comprised three steps:

\begin{enumerate}
    \item \textbf{Bremsstrahlung background modeling.}
    The bremsstrahlung gamma-ray spectrum and its subsequent energy deposition in the LYSO-PX array were modeled using the GEANT4 toolkit (version 11.3) \cite{AGOSTINELLI2003250, 1610988,ALLISON2016186}. The simulation incorporated the full beamline geometry and detector response. For each specific shot, the experimentally measured electron energy spectrum $\mathrm{d}N_e/\mathrm{d}E$ served as the generator input. This yielded a shot-specific, pixelated map of the expected bremsstrahlung background, denoted as $S_{\mathrm{brems}}^{i}$, where the index $i$ runs over all detector pixels.

    \item \textbf{NCS response library generation.}
    The theoretical NCS gamma-ray spectra were calculated using the Locally Monochromatic Approximation (LMA) \cite{Blackburn_2021} for a discrete grid of laser intensities $a_0 \in [1, 10]$. These theoretical spectra were then propagated through the same GEANT4 detector model to generate a response library. This process produced a set of basis vectors, $S_{\mathrm{NCS}}^{i}(a_0)$, representing the expected detector response to a pure NCS signal at a given intensity.

    \item \textbf{Forward-fitting and parameter extraction.}
    The total modeled signal for a given pixel $i$ was constructed as a linear combination of the background and the NCS component: $S_{\mathrm{model}}^{i}(a_0, R) = S_{\mathrm{brems}}^{i} + R \cdot S_{\mathrm{NCS}}^{i}(a_0)$, where $R \in [0, 1]$ is the participant fraction. This two-dimensional model was fitted to the experimental pixel readout $S_{\mathrm{exp}}^{i}$. The best-fit parameters $\hat{a}_0$ and $\hat{R}$ were determined by minimizing the reduced $\chi^2$ statistic across the full detector array. The measurement uncertainty was derived from the $\chi^2$ surface, defined by the contour $\chi^2 = \chi^2_{\mathrm{min}} + 1$ ($1\sigma$ confidence interval). The distribution of the extracted $\hat{a}_0$ values for all valid shots is presented in Fig. 4a.
\end{enumerate}
\subsection*{Gamma-ray polarimetry}
The linear polarization of the generated gamma rays was characterized using two complementary techniques, distinguishing between the lower-energy Compton-dominated regime and the high-energy photonuclear regime.

\begin{enumerate}
    \item \textbf{Compton polarimetry ($< 2.2$ MeV).}
    For the low-energy spectral component, polarization was determined via the azimuthal asymmetry of Compton scattering. The gamma-ray beam impinged on a solid carbon converter target, and the scattered photons were recorded by a downstream Imaging Plate (IP). The azimuthal intensity modulation, $I(\phi) \propto 1 + \mathcal{A}_{\mathrm{C}} P \cos[2(\phi - \phi_0)]$, was fitted to extract the modulation amplitude. Here, $P$ denotes the polarization degree, $\phi_0$ the polarization angle, and $\mathcal{A}_{\mathrm{C}}$ the analyzing power of the detection system. The analyzing power was accurately characterized via GEANT4 simulations (see Fig. 4a), which modeled the differential Compton cross-section and the detector response for fully polarized incident photons.

    \item \textbf{Photoneutron polarimetry ($> 2.2$ MeV).}
    For photons exceeding the deuteron binding energy ($E_{\mathrm{th}} = 2.22~\mathrm{MeV}$), polarization was measured utilizing the $d(\gamma, n)p$ photodisintegration reaction in a heavy-water ($\mathrm{D_2O}$) target. This reaction exhibits a high theoretical analyzing power ($\mathcal{A}_{\mathrm{D}} > 0.9$ for $3 < E_\gamma < 20~\mathrm{MeV}$). The emitted photoneutrons were detected by orthogonal pairs of bubble detectors (BTI-PND, Bubble Technology Industries) aligned parallel and perpendicular to the laser polarization axis.
    To isolate the NCS signal, the neutron background induced by bremsstrahlung was explicitly subtracted. This background contribution was calculated by convolving the bremsstrahlung spectrum (retrieved from the spectral decomposition described above) with the energy-dependent photoneutron yield. The polarization degree was then derived from the measured asymmetry of the background-subtracted neutron doses ($D_\perp, D_\parallel$) via:
    \[
    P = \frac{1}{\bar{\mathcal{A}}_{\mathrm{D}}} \frac{D_\perp - D_\parallel}{D_\perp + D_\parallel},
    \]
    where $\bar{\mathcal{A}}_{\mathrm{D}}$ is the effective analyzing power weighted by the deconvolved NCS spectrum.
\end{enumerate}

    % \subsection{Particle-in-cell simulation}

    % Particle-in-cell(PIC) simulations are performed using the open-source code
    % SMILEI \cite{DEROUILLAT2018351}. The plasma target is 6mm long,consisting of
    % 500$/mu$m linear density up-ramps and down-ramps with a plasma density of $4
    % times 10^{18}$ cm$^{-3}$.A Gaussian laser pulse is focused at 1.2mm from the
    % front edge with $w_{0}$=10.85$\mu$m,and $a_{0}$=4.A SiO$_{2}$ plasma mirror is
    % placed at 6mm with a thickness of 1mm and molecular density of $2.66 times 10
    % ^{22}$ cm$^{-3}$.The simulation box measures 70$mu$m $\times$ 100$m u$m(r
    % $\times$ z) with spatial resolution $\Delta x=2 \Delta y=\lambda/20$.
    \bibliography{sn-bibliography}% common bib file
\end{document}